\documentstyle[twocolumn,epsfig,prl,aps,amsmath,amssymb,float]{revtex}
\begin{document}
\newfloat{figure}{ht}{aux}
\draft
\twocolumn[\hsize\textwidth\columnwidth\hsize\csname
@twocolumnfalse\endcsname
\title{Soliton Approach to Spin-Peierls Antiferromagnets: 
Large-Scale 
Numerical Results}
\author{Erik S\o rensen$^a$, Ian Affleck$^b$, David Augier$^a$ and
Didier Poilblanc$^a$}
\address{$^a$Laboratoire de Physique Quantique \& 
UMR CNRS 5626, Universit\'e Paul Sabatier, 31062 Toulouse, France}  
\address{$^b$Department of Physics and Astronomy, and 
Canadian Institute for Advanced Research, University of British Columbia,\\
Vancouver, BC, V6T 1Z1, Canada}
\date{\today}
\maketitle
\begin{abstract}
A simple intuitive picture of spin-Peierls antiferromagnets arises from
regarding the elementary excitations as S=1/2 solitons. In a strictly
one-dimensional system these excitations are assumed not to form
bound-states and to be repelled by impurities.  
Couplings to the three-dimensional lattice
are assumed to produce an effective confining potential which binds
solitons to antisolitons and to impurities, with the number of
bound-states increasing as the interchain coupling goes to 0.  We
investigate these various assumptions numerically in a phononless model
where spontaneous dimerization arises from frustration and the interchain
coupling is treated in mean field theory. 
\end{abstract}
\vskip2pc]

The discovery of a spontaneous lattice dimerization induced 
by quasi one dimensional antiferromagnetic interactions, the 
spin-Peierls effect, in CuGeO$_3$ has sparked renewed 
experimental
~\cite{hase,isobe,ohama,weiden,regnault,NMR,x-rays,ain,raman,boucher}
and theoretical interest in this field. 
Subjects of current interest include the nature of the excitations 
in the dimerized phase~\cite{uhrig1,uhrig2,brehmer,riera,bouz1},
the effect of impurities~\cite{fukuyama,laukamp,hansen} and the effect of a 
magnetic field~\cite{sakai,uhrig3,schonfeld}.  
A simple intuitive picture of these phenomena is provided by the 
soliton model, first introduced in the context of frustrated spin chains
by Shastry and Sutherland~\cite{shastry}
and discussed for example by Khomskii and 
collaborators~\cite{khomski} as well as others~\cite{brehmer,goff,geilo,hallberg,uhrig4}. 
This model is based on the assumption that 
all interchain interactions, both magnetic
and elastic, are weak. While this assumption may not be very good for
CuGeO$_3$, it is possible that other spin-Peierls systems may be found which
are more one-dimensional.

In the absence of any explicit dimerization a 
completely decoupled single $S=1/2$ chain can still 
have a {\it spontaneously} dimerized 
ground-state due to spin-phonon interactions or due to frustrating next 
nearest neighbor interactions.  The ground-state is two-fold
degenerate, corresponding to the two possible dimerizations, A and B, and 
the fundamental excitations are expected to 
be massive S=1/2 solitons, $s$, and antisolitons, $\bar s$,
with a gap $\Delta_{sol}$, living on different sublattices, 
separating the two different dimerizations.
It is sometimes assumed that no soliton-antisoliton bound-states form in 
this model and that solitons are repelled by non-magnetic impurities, i.e. by 
the ends of an open chain, as we demonstrate below. 
Neither of these assumptions is obvious and indeed 
some approaches make  assumptions in contradiction to these 
ones~\cite{fukuyama}.  

It follows from general principles that spontaneous 
dimerization disappears at any non-zero T, due to a finite 
density of solitons, for a strictly one-dimensional (1D) 
system.  Interchain interactions, either magnetic or 
elastic, can fundamentally change the situation.  
Spontaneous dimerization is now robust against the 
appearance of solitons.  A soliton-antisoliton pair on a 
particular chain separated by some distance $r$ leaves a 
region of the chain in the wrong ground-state relative to 
the neighboring chains.  The consequence is an energy cost 
$\lambda r$ where $\lambda$ is determined by the interchain 
interactions.  Such a linear potential is confining; the 
soliton and antisoliton cannot escape from each other and
behave analogously to quark anti-quark pairs.
Associated with this stabilization of the spontaneously 
dimerized phase is a drastic change in the excitation 
spectrum. As proposed by one of us~\cite{geilo},
low-lying excitations  corresponding
to soliton-antisoliton bound-states now have to occur. These excitations
will have spin $S=1$ or $0$
and energy $E=2\Delta_{sol}+E_b$ where $E_b$ is the binding energy.
There is no soliton continuum due to the confining nature 
of the interaction (i.e. the fact that it grows without 
limit at large separation), however a $s\bar s-s\bar s$ 
continuum can occur beginning at the lowest lying $S=2$ states.
At this energy $E_b$ exceeds $2\Delta_{sol}$ and it becomes
energetically favorable to create a new $s\bar s$ pair. 
Furthermore, a soliton is 
bound by a linear potential to each non-magnetic defect.  
It is convenient to  model the linear potential using a mean field 
treatment of interchain couplings.  In this approach the 
interchain couplings simply provide a term in the 
single-chain
Hamiltonian which favors one or the other of the two 
dimerized states.  It is important to realize that the
soliton picture is only expected to be useful if the 
interchain couplings are relatively small so that the 
solitons, although confined, still behave in a quasi-independent 
fashion at short distances.  If the interchain 
couplings are large then the soliton and antisoliton never 
get far apart compared to their intrinsic size (Compton 
wavelength) and behave effectively as a single well-defined 
magnon. Consequently, the soliton picture ceases to have much utility.

In this paper we make an extensive numerical investigation 
of this soliton picture using the simplest possible model: 
\begin{equation}
H=J\sum_i[(1+\delta(-1)^i){\bf S}_i\cdot{\bf S}_{i+1}+J_2
{\bf S}_i\cdot{\bf S}_{i+1}],
\label{eq:ham}
\end{equation}
a spin only Hamiltonian with spontaneous dimerization 
arising from a second nearest neighbor interaction, $J_2$, and 
explicit dimerization (representing the effect of the 
neighboring chains) due to an alternating nearest 
neighbor interaction, $\delta$.  For $\delta=0$ this model
exhibits a second order phase-transition to a dimerized phase
at $J_{2c}\simeq 0.2411$~\cite{crit}. 
As a representative of the dimerized phase we shall take the
vicinity of the Majumdar-Ghosh (MG) 
model~\cite{MG}, $J_2=\frac{1}{2}$, since at this 
point the correlation length is vanishing and we therefore expect
only minimal finite-size corrections. 
We have performed
exact diagonalizations of up to 32 sites as well as density matrix
renormalization group (DMRG) calculations with $m=128$ states.
Our main results are
as follows: When $\delta=0$
we find that there are no low energy bound-states near zero crystal momentum
for $J_2\leq\frac{1}{2}$. For $J_2>J_{2c}$ the solitons behave as
free massive particles~\cite{us}; we explicitly calculate the gap and
dispersion of the solitons and show 
numerically that the soliton is repelled by the ends of an 
open chain, behaving like a massive particle in a box. 
When $\delta\neq 0$ we demonstrate that a ladder of $S=1,0$
soliton anti-soliton bound-states is formed, increasing in number
with decreasing $\delta$ and giving rise to a range of well-defined peaks
in the dynamical structure factor close to $q=\pi$. 
In this case, an isolated soliton will bind to one of the chain-ends. 
\begin{figure}
\begin{center}
\epsfig{file=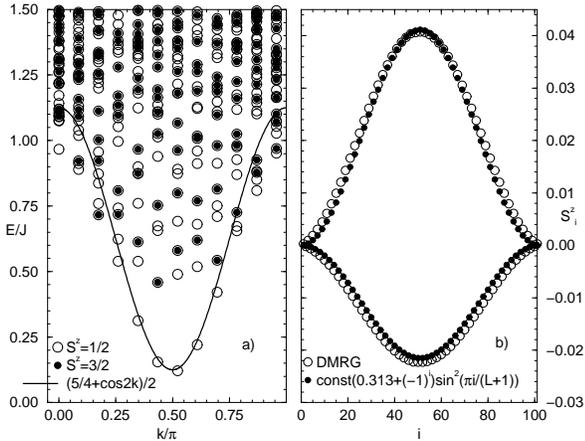,width=7.8cm}
\caption{a) The single soliton dispersion relation at the MG point for
a 23 site chain.
The solid line indicates the Shastry-Sutherland variational estimate
$\omega_{sol}(k)=(J/2)(5/4+\cos 2k )$. b) $<S^z_i>$ as a function of $i$.
DMRG results are shown for L=101, with $S^z_{tot}=1/2,\ m=128$.}
\label{fig:soldisp}
\end{center}
\end{figure}
\begin{figure}
\begin{center}
\epsfig{file=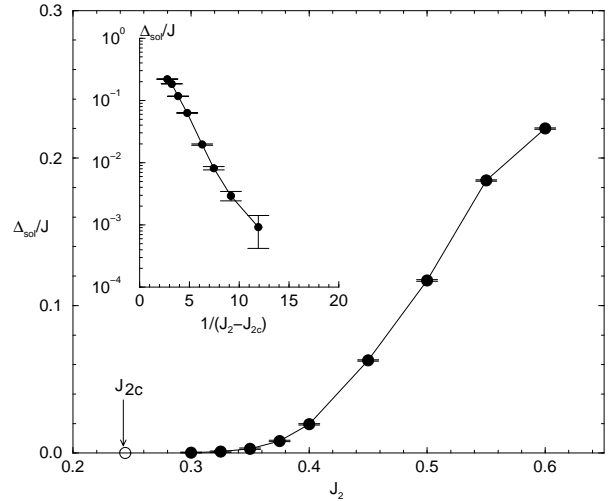,width=7.8cm}
\caption{The soliton gap as a function of $J_2$.}
\label{fig:mass}
\end{center}
\end{figure}
We begin by discussing our results for $\delta=0$.

{\it Odd length systems.}
The behavior of an isolated soliton can be studied by considering
odd-length systems for which the ground-state has $S_{tot}=1/2$. In
Fig.~\ref{fig:soldisp}a we show results for the lowest-lying
$S^z_{\rm tot}=1/2,3/2$ states (open and full circles) 
of a 23 site chain at the MG point. 
The ground-state energy per spin of an even length system, for
$J_2=1/2$ and open boundary conditions, is $-3J/8$ and we use this value
to define zero energy for the odd-site system. A well-defined S=1/2
mode corresponding to the dispersion of the soliton is clearly visible
around $k=\pi /2$. Approximating the soliton as a single unpaired
spin between the two dimer ground-states gives a rigorous upper bound
on the soliton dispersion relation\cite{shastry,arovas} of $E=(J/2)[5/4 + \cos
2k]$. This is shown as the solid line in Fig.~\ref{fig:soldisp}
and agrees very well with the numerical data. Including 3 and 5 spin structures in the
variational soliton wave-function reduces~\cite{cm2} the rigorous
upper bound on the soliton gap, $\Delta_s/J$ (at $k=\pi /2$) from $0.125$
to $0.11701$
in remarkably good agreement with our best DMRG estimate of
0.1170(2)~\cite{us}.  
In Fig.~\ref{fig:soldisp}b $<S^z_i>$ is shown for a $L=101$ site system
in the $S^z_{tot}=1/2$ ground-state.
Clearly the soliton is repelled by the open ends and enters approximately
a particle in
a box state $<S^z_i>\simeq {\rm const}+(-1)^i\sin^2(\pi i/(L+1)$ indicated
by the solid circles. We have calculated the soliton gap defined
as the $\Delta_{sol}=\lim_{L \to\infty} E(L+1)-(E(L)+E(L+2))/2$ 
($L$ even) as a function of
$J_2$. Our DMRG results are shown in Fig.~\ref{fig:mass}. As $J_2$ is
increased from $J_{2c}$ the soliton gap should increase
exponentially~\cite{geilo}: $\Delta_{sol}=\exp(-b/(J_2-J_{2c}))$. 
The numerical data seems largely consistent with such a behavior as can
be seen in the inset of Fig.~\ref{fig:mass} although $\Delta_{sol}$
quickly becomes too small for a reliable determination.

{\it Even length systems.}
DMRG calculations can be performed using spin-inversion as
a symmetry in which case even and odd multiplets can be
distinguished. Performing such calculations for $J_2=1/2$ we find that 
the gap to the lowest-lying triplet and singlet excitations
are degenerate in the thermodynamic limit
$\Delta_{00}/J=\Delta_{01}/J=0.2340(2)$, precisely twice
$\Delta_{sol}$. This degeneracy persists for $J_2<1/2$
and we conclude that 
there are no low energy bound-states near zero crystal momentum, although
such states could possibly occur for $J_2>1/2$.  
On the other hand, at the MG point, Caspers and Magnus (CM)~\cite{cm} have
shown that there are 
exact singlet and triplet bound-state 
at $q=\pi/2$, degenerate
with energy $E/J=1$. The triplet state saturates the total weight of
the dynamical structure factor which is a single delta peak at
$(q=\pi/2,E=J)$. The lowest lying $S=2$ state at $q=\pi/2$
marking the on-set of the continuum has
$E\simeq 1.2J$ and is clearly separated from the bound-state. 
Exact diagonalization results are consistent
with the occurrence of bound-states for a small range of 
momenta close to $q=\pi/2$.  It is interesting to compare this with the 
predictions of the sine-Gordon field theory, expected to be 
valid near $J_{2c}$.
This relativistic field theory, for coupling 
constant, $\beta^2\approx 8\pi$ has no bound-states.  By 
Lorentz invariance, if there are no bound-states at zero 
momentum there cannot be any at finite momentum either.  
However, a resonance with a finite life-time would become 
longer-lived at finite momentum due to relativistic time 
dilation.
Clearly the non 
Lorentz invariance at the MG point is important in allowing 
for bound-states only at finite momentum.  
\begin{figure}
\begin{center}
\epsfig{file=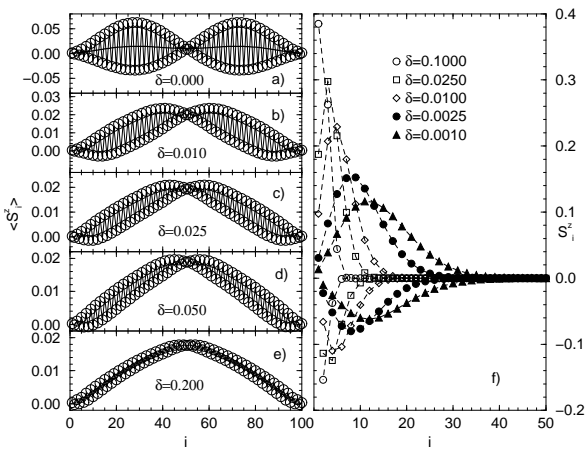,width=7.8cm}
\caption{
a) The on-site magnetization for an {\it even} length chain,
$L=100$, at the MG point, in the two soliton state $S^z_{tot}=1$
and parity $P=-1$.
DMRG results with $m=128$ for different values of the
explicit dimerization $\delta$ is shown. $m=128$ was used in 
the calculation. The solid line, for $\delta=0$, indicates the uniform part
of the magnetization.
f) The on-site magnetization for an {\it odd} length chain,
$L=51$, at the MG point, with $S^z_{tot}=1/2$. 
DMRG results with $m=128$ for different values of the
explicit dimerization $\delta$ is shown. The chain begins with a weak
link.
}
\label{fig:stodd}
\end{center}
\end{figure}
\begin{figure}
\begin{center}
\epsfig{file=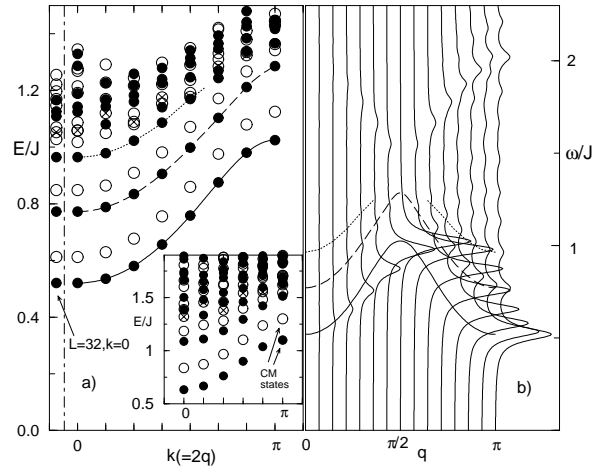,width=7.8cm}
\caption{a) The lowest lying triplet (solid circles), singlets (open
circles) and quintuplets (crosses) for $J_2=0.5$, $\delta=0.05$, $L=28$,
as a function of $k=2q$.
Results for $L=32$, $k=0$ are shown to the left.
The inset shows the same spectrum from $J_2=0.45$, $\delta=0.10$,
$L=20$. b) The dynamical structure factor $S^{zz}(q,\omega)$ 
for $J_2=0.5$, $\delta=0.05$, $L=28$ and a broadening of $\varepsilon=0.04J$.
The solid, dashed and dotted lines indicate the 3 triplet
branches.}
\label{fig:4}
\end{center}
\end{figure}

{\it Explicitly dimerized systems.}
We next investigate the model with an alternating 
interaction, $\delta\neq 0$, added.  From a numerical
perspective two effects complicate
such an investigation; The exponentially diverging correlation
length $\xi$ as $J_{2c}$ is approached and the size,
$l_{s\bar s}$, of a $s\bar s$
bound-state, diverging as $\delta\to 0$. Note that $l_{s\bar s}$
increases with the energy of the bound-state. We need to have
$L\gg \xi,l_{s\bar s}$.
In Fig.~\ref{fig:stodd}a we show DMRG results for the on-site
magnetization $<S^z_i>$ in the lowest-lying two soliton
state $J_2=1/2, S^z_{tot}=1, P=-1, L=100$ for various
different values of $\delta$. For $\delta=0$ two well-defined peaks
can be seen in the uniform part of $<S^z_i>$, separated by $L/2$. As
$\delta$ is increased the formation of a soliton bound-state is
clearly visible and the excitation
becomes more and more magnon like.
In all cases the chain starts and ends with a strong link.
Chains starting and ending with a weak link will have edge like
excitations~\cite{laukamp}.
In Fig.~\ref{fig:stodd}f we show DMRG results for $<S^z_i>$ at
the MG point for different values of $\delta$ as calculated in 
the $S^z_{tot}=1/2$ ground-state. In this case $L=51$ 
is {\it odd} and the first link of the chain is weak. Clearly the
soliton binds to the end of the chain, in agreement with other
recent results~\cite{laukamp}. However, as $\delta$ is decreased
the maximum in $<S^z_i>$ moves further and further inside the chain, only
for the largest value of $\delta$ is the maximum at the first chain
site, as has also been found in the impurity
susceptibility~\cite{laukamp}. From these results we can roughly estimate
the size of a bound-state, $l_{s\bar s}$, as a function of $\delta$. 
We see that we can only hope to determine the
excited states if $J_2\approx 1/2$, if $\delta>0.05$ due to finite-size effects.

We now examine the excitation spectrum for $\delta\neq 0$. In
Fig.~\ref{fig:4}a we show the lowest lying triplet, singlet and
quintuplet excitations vs. $k=2q$ for a system with $L=28, J_2=1/2,
\delta =0.05$. Three well defined triplet branches are clearly visible
below the continuum (the quintuplet $S=2$ states) as well as two singlet
branches. The third and highest lying singlet appears to remain marginally
below the continuum at $k=0$ as can be seen for the $L=32$ results shown
to the left of the panel. The inset in Fig.~\ref{fig:4}a shows the same
spectrum but for $L=20, J_2=0.45, \delta=0.1$. In this case finite size
corrections are significantly smaller and we see that only two triplet 
and two singlet branches are below the continuum. 
We take these results as clear evidence that
the number of bound-states increases as $\delta\to 0$.
At still larger $\delta=0.2, J_2=0.35$ only two triplets and
a one singlet is found~\cite{bouz1}.
The point $J_2=0.45, \delta=0.1$ 
is along the disorder line $\delta=1-2J_2$ where the
excited states of Caspers and Magnus~\cite{cm} remain exact, as
noted by Brehmer et al~\cite{brehmer}, however, they are no longer degenerate.
These two states are indicated 
by an arrow in the inset (CM). As was the case at the MG point the
triplet state saturates the structure factor at $k=\pi$ along the
disorder line, i.e. $S(k,\omega)$ is a single delta peak at the energy
of the triplet.  In Fig.~\ref{fig:4}b we show results for the 
dynamical structure factor $S^{zz}(q,\omega)$ for the $L=28$ results
in panel a. Note the change between $k$ and $q$ from panel a to b,
$k=n\pi/7,\ q=n\pi/14$.
Here,
$S^{zz}(q,\omega)=\sum_n |\langle\Phi_n|S_z(q)|\Phi_0\rangle|^2
\delta_\varepsilon(\omega-E_n+E_0),$
where $E_0$ ($E_n$) is (are) the energy(ies) of the ground-state $\Phi_0$ 
(triplet states $\Phi_n$),  $S_z(q)=\sum_je^{iqr_j}S_j/\sqrt{L}$ is the
Fourier transform of $S_z^j$ and $\delta_\varepsilon$ is a Lorentzian of
width $\varepsilon$. The three triplet branches are most
clearly visible around $q=\pi$. Neutron scattering experiments
on CuGeO$_3$ have so far only seen evidence for a single triplet branch
at $q=\pi$. However, the weight of a peak due to an eventual second triplet
branch should be much smaller and could have been missed. Secondly, it is
possible that other compounds may yield more clear evidence for a ladder
of soliton bound-states which would consolidate the soliton picture.
In conclusion we have demonstrated that well defined soliton
excitations occur in frustrated spin chains. In the absence of
interchain coupling ($\delta=0$) the solitons do not bind
and are repelled by the chain ends. In the presence of
interchain coupling ($\delta\neq 0$) a number
of stable bound-states occurs and isolated solitons are attracted to
the chain ends.
We thank IDRIS (Orsay) for allocation of CPU time on the C94 and C98
CRAY supercomputers. The research of IA is supported in part by NSERC of
Canada.

\end{document}